# MAXWELL DUALITY, LORENTZ INVARIANCE, AND TOPOLOGICAL PHASE

Jonathan P. Dowling* and Colin P. Williams

*Quantum Algorithms & Technologies Group, Mail Stop 126-347*
*Jet Propulsion Laboratory, California Institute of Technology*
*4800 Oak Grove Drive, Pasadena, California 91109*

J. D. Franson

*The Johns Hopkins University*
*Applied Physics Laboratory*
*Laurel, Maryland 20723*

We discuss the Maxwell electromagnetic duality relations between the Aharonov-Bohm, Aharonov-Casher, and He-McKellar-Wilkens topological phases, which allows a unified description of all three phenomena. We also elucidate Lorentz transformations that allow these effects to be understood in an intuitive fashion in the rest frame of the moving quantum particle. Finally, we propose two experimental schemes for measuring the He-McKellar-Wilkens phase.

PACS: 03.65.B, 72.15.R, 14.80.H, 11.30.C

In 1959 Aharonov and Bohm (AB) predicted that a quantum charge, $e = |e|$, circulating around a magnetic flux line would accumulate a quantum topological phase [1], which can be detected using matter-wave interferometry. The flux tube can be thought as a solenoid of infinitesimal cross-sectional area or as a linear array of point magnetic dipoles (Fig. 1a). The AB phase is $\varphi_{AB} = e\Phi_M / \hbar c$, where $\Phi_M = 4\pi^2\mu$ is the magnetic flux with $\mu$ the number of dipoles per unit length. The AB effect has been confirmed by a series of electron interference experiments, culminating in the demonstrations of Tonomura and co-workers [2]. In 1984, Aharonov and Casher (AC) predicted a reciprocal effect [3]. The AC phase accumulates on a quantum, magnetic dipole $\boldsymbol{m}$ as it circulates around and parallel to a straight line of charge (Fig. 1b). The AC phase is given by $\varphi_{AC} = 4\pi m \lambda_E / \hbar c$, where $\lambda_E$ is the electric charge per unit length. It has been observed with a neutron interferometer [4] and in a neutral atomic Ramsey interferometer [5].

He and McKellar in 1993, and Wilkens independently in 1994, predicted the existence of a third topological phase (Fig. 1d) that is essentially the Maxwell dual of the AC effect [6]. We will refer to it here as the He-McKellar-Wilkens (HMW) phase. In the HMW effect, an electric dipole $\boldsymbol{d}$ accumulates a topological phase while circulating around, and parallel to, a line of magnetic charge (monopoles). To relate this to the AC effect, note that Maxwell's equations are invariant under the electric-magnetic duality transformations given by [7], $\mathbf{A}_E \to \mathbf{A}_M$, $\mathbf{E} \to \mathbf{B}$, $e \to g$, $\mathbf{d} \to \mathbf{m}$, and $\mathbf{A}_M \to -\mathbf{A}_E$, $\mathbf{B} \to -\mathbf{E}$, $g \to e$, $\mathbf{m} \to \mathbf{d}$. Here, $g$ is a unit of north magnetic monopole charge, and $\mathbf{A}_E$ and $\mathbf{A}_M$ are electric and magnetic vector potentials, defined in the absence of electric or magnetic monopoles, respectively, such that $\mathbf{B} = \nabla \times \mathbf{A}_B$ if $\nabla \bullet \mathbf{B} = 0$, and $\mathbf{E} = \nabla \times \mathbf{A}_E$ if $\nabla \bullet \mathbf{E} = 0$. As shown by He and McKellar from duality, any derivation of the AC effect is also one of the HMW phase [6]. We can therefore write down the HMW phase as $\varphi_{HMW} = -4\pi d\lambda_M / hc$, by inspection. This result is in agreement with Wilkens's calculation that considered the problem of a dipole moving in an electromagnetic field [6, 8]. Here, $\lambda_M$ is the magnetic monopole charge per unit length, and the minus sign arises from the asymmetric nature of the duality transform. Our duality analysis also predicts a fourth phenomenon, which is the dual of the AB effect (Fig. 1c). Here, a quantum magnetic monopole acquires a topological phase as it circumnavigates a line of electric dipoles (electric flux tube). Any derivation of the Aharonov-Bohm effect is one of the dual Aharonov-Bohm (DAB) effect, through the duality transform. By inspection, the DAB phase is $\varphi_{DAB} = -g\Phi_E / hc$, where $\Phi_E = 4\pi^2\delta$ is the electric flux, with $\delta$ the number of electric dipoles per unit length. All four of these phases are topological in that the result does not depend on the particle velocity or the circulating path taken. They are all nonclassical in that there is no classical force acting on the particle, and that the effect arises in the quantum phase of the wavefunction. In addition, in the AB and the DAB effect, there is the additional feature that there are no $\boldsymbol{B}$ or $\boldsymbol{E}$ fields at the location of the moving charge.

The immediate question is how to observe these dual topological phases experimentally, since they both seem to require isolated magnetic monopoles. For the HMW phase, Wilkens proposed employing a pierced magnetic sheet to mimic the required radial magnetic field and then utilizing a matter-wave interferometer for molecules with a permanent dipole moment [6]. The question of whether or not this scheme will work, even in principle, has been debated [9–11]. So far the effect has not been seen, or at least recognized, in any experiment. It is the purpose of this paper to provide two concrete physical setups in which the HMW phase could be observed and interpreted correctly. Discussion of possible observations of the DAB effect is beyond the scope of this paper.

The DAB, and HMW phases can be written in integral form as

$$\varphi_{DAB} = -\frac{g}{hc}\oint \mathbf{A}_E \cdot \mathbf{dl} = -g\Phi_E / hc, \qquad \varphi_{HMW} = -\frac{1}{hc}\oint [\mathbf{d} \times \mathbf{B}] \cdot \mathbf{dl} = -4\pi m\lambda_M / hc, \qquad (1)$$

where $\mathbf{A}_E$ is the dual-electric vector potential. The AB and AC phases are then obtained immediately from Eqs. (1) by the duality transforms. It is important to note that Hinds and co-workers have shown that the AC effect can be thought of as a motional Zeeman shift [5]. Consider a Lorentz frame $K'$ that is co-moving with any of the circulating particles in Fig. (1). The Lorentz transforms for the electric and magnetic fields are $\mathbf{E}' \cong \mathbf{E} - \mathbf{v} \times \mathbf{B} / c$ and $\mathbf{B}' \cong \mathbf{B} + \mathbf{v} \times \mathbf{E} / c$, in the small velocity limit [7].

Consider the AC effect, Fig. (1b), where $\boldsymbol{B} = 0$ in the lab frame, and $\mathbf{E} = 2\lambda_E \hat{\boldsymbol{\rho}} / \rho$ is the radial electric field from the line of charge. (Here, $\boldsymbol{\rho}$ and $\hat{\boldsymbol{\rho}}$ are the radial and radial-unit vectors, respectively.) As the magnetic dipole circulates about the line of charge, in the co-moving frame the dipole couples to the field via the Hamiltonian, $H'_{AC} = -\mathbf{m} \cdot \mathbf{B}'$, giving rise to a Zeeman phase shift, $\varphi_{AC} = \oint H'(t) dt / \hbar$, integrated over the circulation time. This is precisely the AC phase when one Lorentz transforms back into the lab frame and performs a change of variable, $\boldsymbol{v} = d\boldsymbol{l}/dt$. Our argument here is good to first order in $v/c$, whereas the more complete treatment of Hinds and colleagues is good to all orders [5]. Maxwell duality tells us immediately that the dual HMW phase is then—to all orders in $v/c$—a motional *Stark* effect, with $H'_{HMW} = -\mathbf{d} \cdot \mathbf{E}'$ in the co-moving frame of the electric dipole. In the HMW lab frame, $\boldsymbol{E} = 0$, and hence $H_{HMW} = \mathbf{d} \cdot (\mathbf{v} \times \mathbf{B} / c)$, which is the Röntgen interaction employed in the derivation by Wilkens [6,8].

For completeness, let us also analyze the AB and DAB effects in the co-moving frame. In the small velocity limit, the potentials transform as the components of a covariant four-vector [7], $V' \cong V - \mathbf{v} \cdot \mathbf{A} / c$, $\mathbf{A}'_{||} \cong \mathbf{A}_{||} - V\mathbf{v} / c$, and $\mathbf{A}'_{\perp} \cong \mathbf{A}_{\perp}$. Here, $V$ is the scalar potential, and the parallel and perpendicular notation is with respect to the direction of motion. The general interaction Hamiltonian is $H_{AB} = -eV + e\mathbf{p} \cdot \mathbf{A} / mc$, but this reduces to $H'_{AB} = -eV'$ in the co-moving frame where the momentum $\boldsymbol{p}' = 0$. Lorentz transforming to the lab frame, we get,

$$\varphi_{AB} = \frac{1}{\hbar} \oint H'_{AB}(t) dt = -\frac{e}{\hbar} \oint V' dt = \frac{e}{\hbar c} \oint \mathbf{A} \cdot \mathbf{v} dt = \frac{e}{\hbar c} \oint \mathbf{A} \cdot \mathbf{dl} , \qquad (2)$$

which is the AB phase, Eq. (1). The interpretation is that in the co-moving frame the circulating charge experiences a constant voltage difference across the branches of the interferometer. This electrostatic potential gives rise to the AB phase shift when integrated over the circulation time. It is easy to show that $\boldsymbol{E}'$ and $\boldsymbol{B}'$ are still zero. Similar arguments hold for the DAB phase.

We move to our experimental proposals for observing and interpreting the HMW phase. The key is to employ the same transformation that Hinds and co-workers used to enhance the AC effect in neutral atoms [5]. The difficulty with a demonstration of the AC effect with matter-wave interferometry is that it is challenging to maintain a large enough voltage on a single wire of charge to see the effect in the configuration of Fig. (1b). In the inset of Fig. (2), one sees that the same AC phase would be obtained if the dipoles were made to follow identical paths on one side of the wire, but in a superposition of up and down magnetic moments. Then the single wire may be replaced with a parallel-plate capacitor that maintains the same constant electric field along a now straight trajectory, inset of Fig. (2). Since the electric field across the two plates can be made very large compared to that near an isolated wire, this enhances the AC for easy measurement. The superposition of magnetic moments is prepared by exciting an ensemble of magnetic sublevels, and the measurement of the phase is made by two-pulse Ramsey interferometry [5].

By Maxwell duality, the same transformation will work for a measurement of the HMW effect in neutral atoms. In this case the plus and minus signs in the inset of Fig. (2) represent magnetic charge, and the moving quantum dipole is electric. This amounts to replacing the unphysical line of magnetic monopoles with the north and south poles of an ordinary magnet, and the need for the line of magnetic monopoles has been eliminated. Finally, we need only specify an atomic state that can be prepared in a superposition of up and down electric dipole moments. Nature provides us with such systems in excited hydrogen-like atoms and ground-state ammonia molecules .

Consider the original HMW configuration of Wilkens transformed into the Hinds configuration of the inset of Fig. (2), which requires a superposition of dipoles in one atom on one path. The excited hydrogen-like atoms have a degeneracy of states with opposite parity. Hence, the superposition of such levels results in energy eigenstates without definite parity. That is, the expectation of their electric dipole moment does not vanish [12]. This effect is responsible for the first-order Stark effect in hydrogenic atoms and vanishes in non-hydrogenic atomic systems. Consider the *2s* and *2p* first-excited states of hydrogen, $|nlm\rangle = |2lm\rangle$, where $l \in \{0, 1\}$ and $m \in \{-1, 0, 1\}$. Applying degenerate perturbation theory, the linear Stark effect splits the degenerate $m = 0$ level into two components, with energy eigenvalues and eigenfunctions given by $\Delta\varepsilon_\pm = \pm 3 a_B e E'$ and $|\psi^\pm\rangle = \{|200\rangle \pm |210\rangle\}/\sqrt{2}$, respectively [12]. Here, $E'$ is the electric field in the co-moving frame. The $|nlm\rangle$ eigenstates have definite parity and hence zero dipole expectation value. However, in the new basis the states $|\psi^\pm\rangle$ are not parity eigenstates and have definite up and down dipole values. Inverting the transformation, we find for the *2s* and *2p* states $|200\rangle = \{|\psi^+\rangle + |\psi^-\rangle\}/\sqrt{2}$ and $|210\rangle = \{|\psi^+\rangle - |\psi^-\rangle\}/\sqrt{2}$, respectively. This representation shows that these excited states are equal superpositions of up and down electric dipole moments— the condition that is required for the Hinds-transformed HMW configuration. (For this argument, we have suppressed the hyperfine structure dependence, since the bulk of the effect is due to the degeneracy of the orbital magnetic sublevels. Hyperfine effects are restored in the exact calculation below.)

For our first proposed experiment we can use a simple time-of-flight measurement. Suppose we have a source of metastable hydrogen atoms in the *2s* state (Fig. 2), which are easily made [13–16, 18]. Such atoms have an exceedingly long lifetime of about $\tau_{2s} = 0.14$ s, since the transition to the ground state is dipole forbidden [17]. At typical atomic beam velocities of $v = 10^6$ cm/s, the lifetime is effectively infinite during passage through a *1.0* cm magnetic field region. As the atom propagates the metastable *2s* state will, in the co-moving frame, experience a field $E'$ and will be Stark shifted. Integrating over the time of flight between the magnet poles and transforming into the lab frame, the HMW phase and the time-dependent wavefunction become [17],

$$\varphi_{HMW} = 3a_B eBL / \hbar c = 3a_B eBvt_0 / \hbar c \ , \tag{3a}$$

$$|\Psi(t)\rangle = |200\rangle \cos(\varphi_{HMW}\, t/t_0) + i|210\rangle \sin(\varphi_{HMW}\, t/t_0) \ . \tag{3b}$$

Here, $a_B$ is the hydrogenic Bohr radius, $L$ the distance flown between the magnetic poles, and $t_0$ the corresponding time of flight. If $B$ is measured in Gauss and $L$ in centimeters, then Eq. (3a) can be written $\varphi_{HMW} = 0.24\ BL$. However, the *2p* state decays rapidly to the ground state by a dipole transition and has a very short lifetime of only $\tau_{2p} = 1.6 \times 10^{-9}$ s. Therefore, once in the *2p* state the atom will decay into *1s* over a distance of about 16 mm; long before reaching the detector. As the wavefunction oscillates between *2s* and *2p*, as per Eq. (3b), the HMW phase shift converts metastable *2s* atoms into ground state *1s* hydrogen at a rate depending on the phase-dependent oscillation period, $T = \pi t_0 / \varphi_{HMW}$. This electric-field-induced decay is called the "Stark quenching" of metastable, hydrogen-like atoms. It is a previously measured effect that has been seen with both applied external electric fields [14] and with magnetic-induced motional Röntgen electric fields [15,16,18]. In the latter case, the effect was used as a magnetic-field dependent beam polarizer in 1952 by Lamb and Retherford [15] and more recently by Robert, *et al.,* as a velocity selector [16]. The earliest observation of motional Stark quenching was probably made in 1916 by Wien [18]. What is new here is the interpretation of the phenomenon in terms of the HMW phase.

The dumping of *2s* to *1s* states becomes very efficient when the period of the *2s* to *2p* oscillation is nearly equal to the *2p* decay rate [17]. This resonance occurs at the motional Stark-induced level crossing between *2s* and *2p* [16]. We will assume a beam velocity of $v = 10^6$ cm/s. In this case, the resonance condition requires a rather strong field of $B = 8.12 \times 10^3$ G, which will induce almost complete conversion. However, at smaller fields we may take the full motional-Stark-quenched, radiation-rate equations [14], and integrate them over solid angle to compute the total field-induce atomic decay rate $\gamma$. Restoring the suppressed hyperfine dependence, to first order in $B$ and the fine structure constant $\alpha$, this rate can be written as,

$$\gamma(2s_{1/2} \rightarrow 1s_{1/2}) = \frac{3^{10}}{2^8} \frac{1}{\alpha^3} \frac{a_B^3}{\hbar} \left(\frac{vB}{c}\right)^2 = \frac{3^8}{2^8} \frac{1}{\alpha^4} \frac{a_B}{ct_0^2} \varphi_{WR}^2 \ , \tag{4}$$

where the HMW phase $\varphi_{HMW}$ is given by Eq. (3a). If we take $B$ in units of Gauss, this can be written conveniently as $\gamma = 92.6\ B^2$, emphasizing the quadratic dependence on the magnetic field. Taking an exponential with $L = 1.0$ cm, then $t_0 = 10^{-6}$ s and the initial metastable flux in the $2s_{1/2}$ state will decay to *1/e* of its original value for a field on the order of $B \cong 100$ G. These numbers are in good agreement with the velocity-selection experiments of J. Robert, *et al*. [16]. Such a magnetic field induces a HMW phase shift of around $\varphi_{iHMW} \cong 8\pi$. Our calculation includes all relativistic effects, including the hyperfine spin structure, to first order in $\alpha$. We start with the full QED expression of Drake, *et al.,* for the Stark-quenched photon angular decay probability, averaged over the atomic spin-polarization [14]. We then average the photon flux over spherical angle to get the total emission rate, which we identify with the atomic decay rate.

For a second experimental consideration, Chiao has pointed out that the ammonia molecule can be prepared in a superposition of up and down electric dipoles [19]. This can be done by applying a $\pi/2$ microwave pulse at the $NH_3$ ground vibrational two-level inversion splitting at $\omega = 23$ GHz. The result is an equal superposition of the two inversion states, which have opposite parity. Thus the state has no definite parity and hence a nonzero dipole moment, which is susceptible to the motional Stark effect. The HWM phase becomes $\varphi_{HMW} = d_a BL / \hbar c$, where $d_a$ is now the ammonia dipole moment. However, both inversion states of the ammonia molecule are stable, and hence one has time to use two-pulse Ramsey spectroscopy on the system, as was done in 1951 with the first ammonia molecular clock experiments of Lyons, *et al.* [20]. The set up would be the same as in Fig. (2), but now with an ammonia beam and $\pi/2$ microwave pulses applied at both the entrance and exit of the magnetic field region. The HMW phase is lifted directly out of the Ramsey interference fringes in a state-dependent detection process, analogous to the measurement AC effect in neutral atoms [5]. The phase shift should be easily visible with magnetic fields on the order of 100 Gauss.

In summary, we have considered the Maxwell duality transformations among the four topological phases found in the Aharonov-Bohm (AB), Aharonov-Casher (AC), dual Aharonov-Bohm (DAB), and He-McKellar-Wilkens (HMW) effects. In particular, the DAB and HMW effects are derived trivially from the AB and AC effects via duality. In addition, we have looked at the simplification that comes from Lorentz boosting into the co-moving frame of the quantum particle. In this frame the AB and DAB fields appear as if induced by a static potential differential, and the AC and HMW effects are interpreted as motional Zeeman and Stark effects, respectively. Finally, we propose a specific transformation of the HMW configuration into an ordinary dipole electromagnet set up, that allows for the experimental observation and interpretation of the HMW phase. This phase can be seen in the excited states of hydrogen-like atoms, via the first order Stark shift in the co-moving frame. Such motional Stark shifts have been seen already experimentally, but have not hitherto been interpreted in terms of the HMW phase [15, 16, 18]. We also propose a different direct HMW phase measurement experiment using a Ramsey two-pulse interferometer with ammonia beams and microwaves [20].

## ACKNOWLEDGEMENTS

We would like to acknowledge interesting and useful discussions with Y. Aharonov, R. Y. Chiao, X. G. He, E. A. Hinds, L. Maleki, B. H. J. McKellar, T. B. Pittman, and M. Wilkens. A portion of this work was carried out under a contract with National Aeronautics and Space Administration. Additional support was provided by the Office of Naval Research.

**FIGURE CAPTIONS**

Fig. 1    We indicate the Aharonov-Bohm (AB), Aharonov-Casher (AC), dual AB (DAB), and He-McKellar-Wilkens (HMW) topological phase configurations in (a), (b), (c), and (d). Here, $e$ is an electric charge, $g$ is a magnetic (monopole) charge, $m$ is a point magnetic dipole and $d$ is an electric dipole.

Fig. 2    Experiment to measure the HMW phase. Metastable hydrogen atoms are generated in the atom source, and then enter the magnetic field region where the HMW phase couples the metastable *2s* to the rapidly decaying *2p.* The excited state experiences a HMW phase shift that is proportional to magnetic field *B,* producing a phase-dependent dumping of the metastable *2s* into the ground state *1s,* which is then detected. Inset: Starting with the original AC configuration (left), note that the same effect occurs if the particle circulates only counter clockwise in a superposition of up and down dipoles, (middle). Since the field is constant along the path, we replace the line of charge with a parallel plate configuration (right). This same transformation applies to the dual HMW effect, converting a line of magnetic monopoles in to an ordinary dipole magnet.

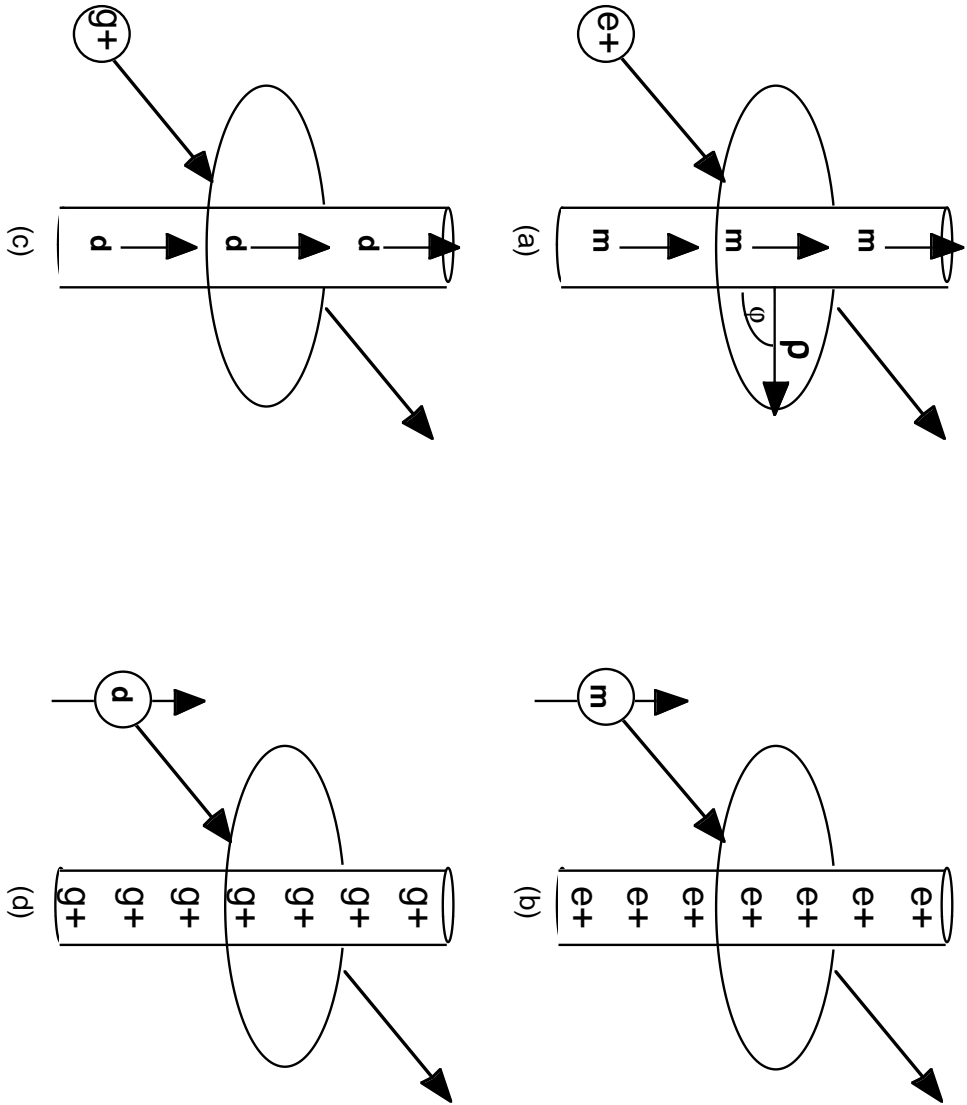

Fig. 1

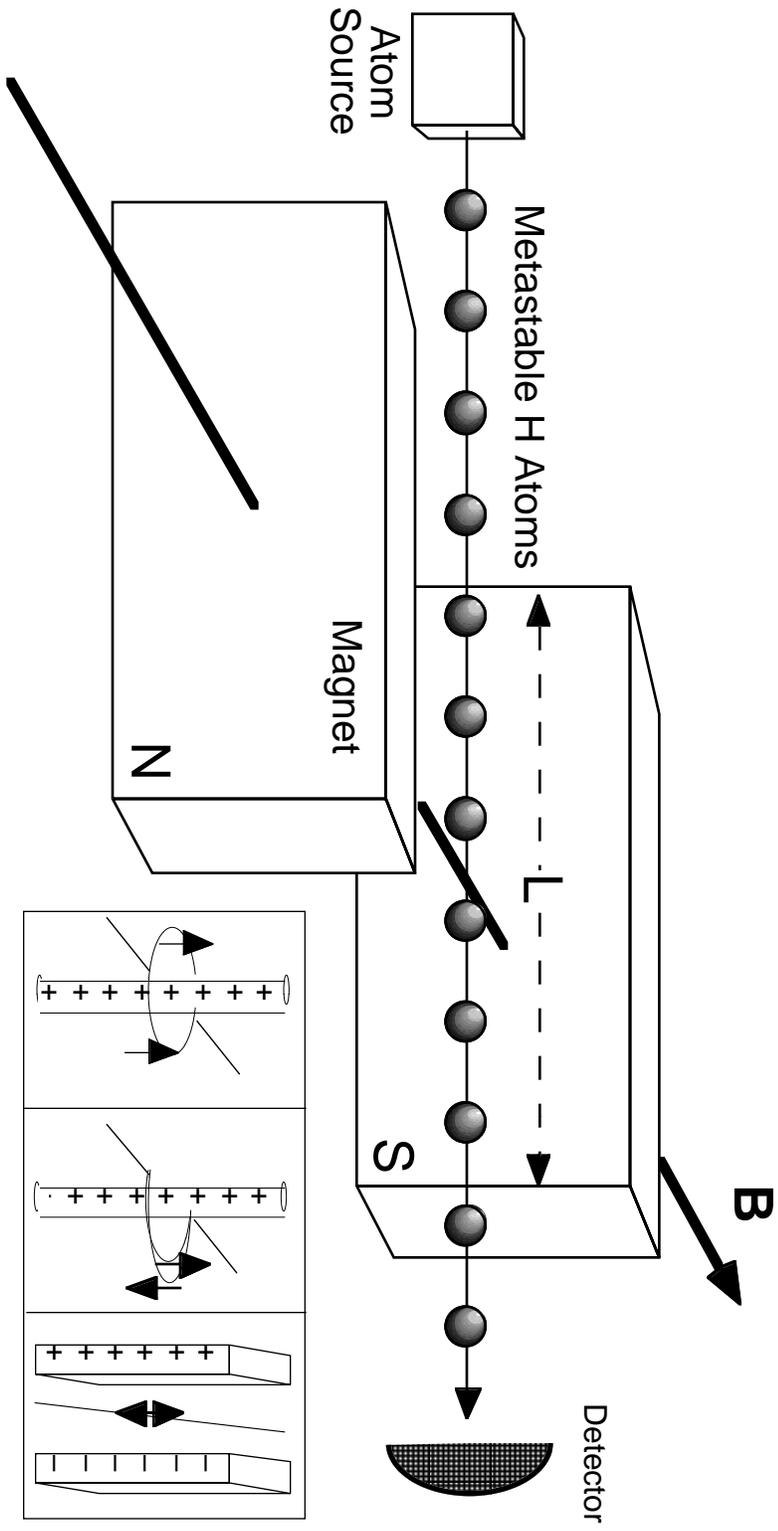

Atom Source

Metastable H Atoms

Magnet

N

S

L

B

Detector

Fig. 2